\documentclass[12pt]{article}
\usepackage{bbm}
\newcommand{\be}{\begin{equation}}
\newcommand{\ee}{\end{equation}}
\newcommand{\ba}{\begin{array}}
\newcommand{\ea}{\end{array}}

\newcommand{\hs}{\mathbf{H}}
\begin{document}
\title{\normalsize \hfill UWThPh-2006-1\\[1cm]
\LARGE
An open--quantum--system formulation\\
of particle decay}
\author{
Reinhold A. Bertlmann,\thanks{E-mail: reinhold.bertlmann@univie.ac.at}\, Walter
Grimus,\thanks{E-mail: walter.grimus@univie.ac.at}\, and Beatrix C.
Hiesmayr\thanks{E-mail: beatrix.hiesmayr@univie.ac.at}
\\[2mm]
\small Institut f\"ur Theoretische Physik, Universit\"at Wien \\
\small Boltzmanngasse 5, A--1090 Wien, Austria \\[3.6mm]}

\date{February 14, 2006}

\maketitle

\begin{abstract}
We consider an open quantum system which contains unstable states. The time evolution of
the system can be described by an effective non-hermitian Hamiltonian $H_\mathrm{eff}$,
in accord with the Wigner--Weisskopf approximation, and an additional term of the
Lindblad form, the socalled dissipator. We show that, after enlarging the original
Hilbert space by states which represent the decay products of the unstable states, the
non-hermitian part of $H_\mathrm{eff}$ ---the ``particle decay''--- can be incorporated
into the dissipator of the enlarged space via a specific Lindblad operator. Thus the new
formulation of the time evolution on the enlarged space has a hermitian Hamiltonian and
is probability conserving. The equivalence of the new formulation with the original one
demonstrates that the time evolution which is governed by a \textit{non-hermitian}
Hamiltonian and a dissipator of the Lindblad form is nevertheless completely positive,
just as systems with hermitian Hamiltonians.
\\[2mm]
PACS numbers:
03.65.-w, 03.65.Yz, 13.25.Es \\[2mm]
Key-Words: Open quantum system, Lindblad operator, particle decay,
Wigner--Weisskopf approximation
\end{abstract}

\newpage

\paragraph{Introduction:}
In reality a quantum system is not isolated but always interacting with its environment and has to
be considered as an open quantum system~\cite{BreuerPetruccione,AlickiFannes}. It leads to a
mixing of the states in the system--- decoherence---and to an energy exchange between system and
environment---dissipation~\cite{Joos,KublerZeh,JoosZeh,Zurek}. The decoherence/dissipation weakens
or destroys the typical quantum phenomena, the interferences.

There is great interest to study such decoherence/dissipation processes for elementary particles
like the ``strange'' K-mesons (neutral kaons) and the ``beauty'' B-mesons
\cite{EllisHagelinNS,HuetPeskin,EllisLopezMN}, in particular, the decoherence of entangled meson
pairs~\cite{BertlmannDurstbergerHiesmayr2002,BG3,BernabeuMavromatosP} (for an overview see Ref.
\cite{BertlmannSchladming}). However, these particles are decaying, which leads in the framework
of quantum mechanics to some difficulties. In the usual framework the decaying states evolve
according to the Wigner--Weisskopf approximation (WWA) where the probability of detecting the
particle is not conserved and the corresponding Hamiltonian is, for this reason, non-hermitian.

The master equation for such an open quantum system described by the density
matrix $\rho(t)$ is given by
\begin{equation}\label{Lindbladequation}
\frac{\mathrm{d}\rho}{\mathrm{d}t} =
-\,i H_\mathrm{eff} \rho \,+\,i\rho H_\mathrm{eff}^\dagger\,-\,D[\rho] \,,
\end{equation}
where the dissipator has the following general structure of
operators~\cite{Lindblad,GoriniKossakowskiSudarshan}:
\begin{equation}\label{dissipator}
D[\rho] = \frac{1}{2} \,\sum_j \left(
A^\dagger_j A_j \,\rho + \rho A^\dagger_j A_j
\,-\, 2 A_j \rho A_j^\dagger \right).
\end{equation}
The operators $A_j$ are usually called Lindblad operators.
The effective Hamiltonian of the system is given, according to the WWA,
by
\begin{equation}\label{hamiltonian}
H_\mathrm{eff} \; = \; H - \frac{i}{2} \,\Gamma,
\end{equation}
where $H$ and $\Gamma$ are both hermitian and, in addition, $\Gamma \geq 0$.

In terms of open quantum systems, the time evolution of the density matrix
represents a
dynamical map which transform initial density matrices $\rho(0)$ on the
Hilbert space of
states, $\hs_s$, to final density matrices $\rho(t) \,=\, V_t[\rho(0)]$, while
the system
is interacting with an environment. Such dynamical maps are
i) trace conserving,
ii) convex linear,
iii) completely positive.
Complete positivity is a rather strong and
important property. It is defined by demanding that all extensions
$V_t \otimes \mathbbm{1}_n$ on the Hilbert space
$\hs_s \otimes \mathbbm{C}^{n}$ are positive, i.e.
\begin{equation}
(V_t \otimes \mathbbm{1}_n)[\rho] \geq 0 \quad
\forall \,t \geq 0, \;\,\forall \,n=0,1,2,\dots,
\;\, \mbox{and}\;\,
\forall \,\rho \geq 0,
\end{equation}
where $\rho$ is a density matrix on $\hs_s \otimes \mathbbm{C}^{n}$.
Physically, it is a reasonable condition since the extension
$V_t \otimes \mathbbm{1}_n$ can be considered in quantum information as an
operator on the composite quantum system
Alice and Bob, which acts locally on Alice's system without influencing Bob.
Furthermore,
complete positivity essentially ensures that tensor products of maps $V_t$
remain positive, an important property especially when considering entangled
states.

However, extracting from a dynamical map the Hamiltonian term
in the usual way~\cite{Preskill-notes} leads to a hermitian Hamiltonian in a
master equation like Eq.~(\ref{Lindbladequation}).
In this sense the authors of Ref.~\cite{Caban-etal}
investigated recently decoherence \textit{and} decay of the neutral kaons by
using a dynamical map corresponding to a master equation with hermitian
Hamiltonian. Nevertheless,
many authors work with a non--hermitian Hamiltonian---which is a standard
procedure in particle physics (see, e.g., Ref.~\cite{QuangPham})---in order to
include the decay property of the particle.

\paragraph{The formalism:}
We are going to show in this letter that both approaches---each one having
its own appeal---are indeed equivalent to each other for quite general
quantum systems. To be more precise, we ask the following question:\\
\textit{Can we work with a hermitian Hamiltonian and incorporate the decay as Lindblad operator
such that the time evolution of the system represents a completely positive map? Moreover, does it
describe the WWA properly without effecting decoherence and/or dissipation?}\\
The answer will be yes, it is possible! But we have to enlarge the Hilbert
space and include formally the decay products of the unstable states.

Thus we assume that the total Hilbert space
$\hs_\mathrm{tot}$ is the direct sum
\begin{equation}\label{Hilbertspace}
\hs_\mathrm{tot} = \hs_s \oplus \hs_f,
\end{equation}
where $\hs_s$ contains the states of the system we are interested in whereas
$\hs_f$ is the space of the ``decay states,''
defined in the following way. We assume that $d_s
\equiv \dim \hs_s < \infty$. Then the non-hermitian part $\Gamma$ of the
effective Hamiltonian in Eq.~(\ref{hamiltonian}) can be decomposed as
\begin{equation}\label{gamma}
\Gamma = \sum_{j=1}^r \gamma_j\, |\varphi_j \rangle \langle \varphi_j |
\quad \mbox{with} \quad \gamma_j > 0 \;\,\forall\, j,
\end{equation}
with an orthonormal system $\{ \varphi_j \}$ in $\hs_s$ and $r = \dim \hs_s - n_0$, where $n_0$
denotes the degeneracy of the eigenvalue zero of $\Gamma$. We demand that $\dim \hs_f \equiv d_f
\geq r$, otherwise $\hs_f$ is arbitrary.

In the following, $\rho$ denotes a density matrix on $\hs_\mathrm{tot}$,
where it has the following decomposition:
\begin{equation}\label{densitymatrix}
\rho =
\left( \ba{cc} \rho_{ss} & \rho_{sf} \\
\rho_{fs} & \rho_{ff} \ea \right) \qquad \mathrm{with} \qquad
\rho_{ss}^\dagger = \rho_{ss}, \quad
\rho_{ff}^\dagger = \rho_{ff}, \quad
\rho_{fs}^\dagger = \rho_{sf}.
\end{equation}
Now we have to define the time evolution $\rho$ on $\hs_\mathrm{tot}$. The
Hamiltonian $H$ and the Lindblad operators $A_j$ are easily extended to
the total Hilbert space by
\begin{equation}\label{operators-H-A}
\mathcal{H} = \left( \ba{cc} H & 0 \\ 0 & 0 \ea \right), \qquad
\mathcal{A}_j =
\left( \ba{cc} A_j & 0 \\ 0 & 0 \ea \right) \,.
\end{equation}
Note that for the definition of $\mathcal{H}$ we have used the
\emph{hermitian} part $H$ of $H_\mathrm{eff}$.

Now we turn to the decay. We need a Lindblad operator $\mathcal{B}$ on the
full space $\hs_\mathrm{tot}$ which describes the decay in the subspace
$\mathcal{H}_s$.
As we will see in the next paragraph, the decay is described by
\begin{equation}\label{operator-B}
\mathcal{B} = \left( \ba{cc} 0 & 0 \\ B & 0 \ea \right)
\quad \mbox{with} \quad
B: \, \hs_s \to \hs_f
\end{equation}
and
\begin{equation}\label{BB}
\Gamma = B^\dagger B.
\end{equation}
Let $\{ f_k \}$ be an orthonormal basis of $\hs_f$. Then we can
decompose $B$ as
\begin{equation}
B = \sum_{k=1}^{d_f}\, \sum_{j=1}^r b_{kj} \,
| f_k \rangle \langle \varphi_j |.
\end{equation}
In order to fulfill Eq.~(\ref{BB}), we require
\begin{equation}
\sum_{k=1}^{d_f} b_{ki}^* b_{kj} = \delta_{ij}\, \gamma_j.
\end{equation}
For $d_f \geq r$, such a $d_f \times r$ matrix $\left( b_{kj} \right)$
always exists. The simplest case is $d_f = r$, where we can choose
\begin{equation}
B = \sum_{j=1}^{d_f} \sqrt{\gamma_j}\,
| f_j \rangle \langle \varphi_j |.
\end{equation}
In that case, each unstable decaying state $| \varphi_j \rangle$ would decay
into just one specific decay state $| f_j \rangle$.

Then we can write the following master equation for the density matrix on
$\hs_\mathrm{tot}$:
\begin{equation}\label{Lindbladequation-onHtot}
\frac{\mathrm{d}\rho}{\mathrm{d}t} =
-\,i \,\big [\mathcal{H},\rho \big ] \,-\,\mathcal{D}[\rho],
\end{equation}
with the dissipator
\begin{equation}\label{dissipator-onHtot}
\mathcal{D}[\rho] = \frac{1}{2} \,\sum_j
\left( \mathcal{A}^\dagger_j \mathcal{A}_j\,
\rho \,+\, \rho \,\mathcal{A}^\dagger_j \mathcal{A}_j \,-\, 2 \,\mathcal{A}_j
\rho \mathcal{A}_j^\dagger \right) +
\frac{1}{2} \, \left( \mathcal{B}^\dagger
\mathcal{B}
\,\rho \,+\, \rho \,\mathcal{B}^\dagger \mathcal{B} \,-\, 2 \,\mathcal{B} \rho
\mathcal{B}^\dagger \right).
\end{equation}
To prove that this master equation contains Eq.~(\ref{Lindbladequation}),
we decompose it with respect to the components of $\rho$:
\begin{eqnarray}
\dot{\rho}_{ss} & = & -i \,\big[ H,\rho_{ss} \big] \,-\, \frac{1}{2}
\,\big\{
B^\dagger B, \rho_{ss} \big\} \,-\, \frac{1}{2} \sum_j \left( A_j^\dagger A_j
  \,\rho_{ss}
+ \rho_{ss} A_j^\dagger A_j - 2\, A_j \rho_{ss} A_j^\dagger \right),
\label{ss} \\
\dot \rho_{sf} & = & -i \,H \rho_{sf} \,-\, \frac{1}{2} \,B^\dagger B
\,\rho_{sf}
\,-\, \frac{1}{2} \sum_j A_j^\dagger A_j \,\rho_{sf} \,,
\label{sf} \\
\dot \rho_{ff} & = & B \rho_{ss} B^\dagger. \label{ff}
\end{eqnarray}
Indeed, with Eq.~(\ref{BB}), we immediately see that Eq.~(\ref{ss}) for $\rho_{ss}$ reproduces the
original equation~(\ref{Lindbladequation}). Furthermore, it is obvious from
Eqs.~(\ref{Lindbladequation-onHtot}) and (\ref{dissipator-onHtot}) that $\mbox{Tr}\, \rho(t) = 1
\quad \forall\, t \geq 0\,$.

At this point, some general remarks are at order. By construction, the time evolution of
$\rho_{ss}$ is independent of $\rho_{sf}$, $\rho_{fs}$, and $\rho_{ff}$. Actually, the time
evolution of $\rho_{sf}$ or $\rho_{fs}$ completely decouples from that of $\rho_{ss}$ and
$\rho_{ff}$. The time evolution of $\rho_{ff}$, the density matrix of the decay states, is
determined solely by $\rho_{ss}(t)$ and is the characteristic for a decay process!

With the initial condition $\rho_{ff}(0) = 0$, it is simply given by
\begin{equation}\label{rho-ff}
\rho_{ff} (t) = B \int_0^t \mathrm{d}t' \rho_{ss}(t') \,B^\dagger.
\end{equation}
If we choose the initial condition
$\rho_{sf}(0) = 0$, then $\rho_{sf}$ remains zero for all times and the
same applies to $\rho_{fs}$. Anyway, these parts of $\rho$ are
totally irrelevant for our discussion.

Generally, for decaying systems the following properties of the time
evolution~(\ref{Lindbladequation-onHtot}) are most important:
\begin{itemize}
\item
$\rho_{ss}(t)$ and $\rho_{ff}(t)$ are positive $\forall\,t \geq 0$;
\item
The time evolution of $\rho_{ss}$ is \textit{completely positive}.
\end{itemize}
\textbf{Proof:} The density matrix $\rho(t)$ obeys the time evolution given by the master
equation~(\ref{Lindbladequation-onHtot}), therefore, $\rho(t)\geq 0$ $\forall \,t \geq 0$. Thus,
confining ourselves to vectors $v \in \mathcal{H}_s$, we evidently have $0 \leq \langle v |
\rho(t) v \rangle \equiv \langle v | \rho_{ss}(t) v \rangle$ and, therefore, $\rho_{ss}(t) \geq
0$. The same reasoning holds for $\rho_{ff}(t)$. As for complete positivity, we note that on
$\hs_\mathrm{tot} \otimes \mathbbm{C}^n$ a general operator has the structure $R = \sum_k \sigma_k
\otimes L_k$, with $\sigma_k$ operating on $\hs_\mathrm{tot}$ and $L_k$ on $\mathbbm{C}^n$. We
know that the dynamical map $V_t$ induced by Eq.~(\ref{Lindbladequation-onHtot}) is completely
positive. Therefore, if $R \geq 0$, then we have $(V_t \otimes \mathbbm{1}_n)[R] \geq 0$. We also
know from Eqs.~(\ref{ss}), (\ref{sf}), and (\ref{ff}), that $V_t$ has a well defined restriction
to operators acting purely on $\hs_s$. Thus, $R \geq 0$ on $\hs_s \otimes \mathbbm{C}^n$ is mapped
into a positive operator on the same space by $V_t \otimes \mathbbm{1}_n$. This concludes the
proof.

\paragraph{The case of a non-singular $\Gamma$:}
Finally, it is interesting to consider the special case of a non-singular
$\Gamma$, where all states of $\hs_s$ decay. Then,
$\dim \hs_f \geq \dim \hs_s$ and, in addition to what we discussed before,
the following properties of the
time evolution~(\ref{Lindbladequation-onHtot}) hold:
\begin{enumerate}
\renewcommand{\labelenumi}{\roman{enumi})}
    \item $\lim_{t \to \infty} \rho_{sf}(t) = 0$,
    \item $\lim_{t \to \infty} \rho_{ss}(t) = 0$,
    \item $\lim_{t \to \infty} \mathrm{Tr}\, \rho_{ff}(t) = 1$.
\end{enumerate}
\textbf{Proof:}
\begin{enumerate}
\renewcommand{\labelenumi}{\roman{enumi})}
\item
We consider the equation
\begin{equation}\label{omega}
\frac{\mathrm{d}}{\mathrm{d}t} \,\omega =
-iH \,\omega \,-\, \frac{1}{2} \,\Gamma' \,\omega \,,
\quad \mbox{with} \quad
\Gamma' = \Gamma + \sum_j A^\dagger_j A_j
\end{equation}
and $\omega(t) \in \hs_s$. From Eq.~(\ref{omega}) we derive
\begin{equation}
\frac{\mathrm{d}}{\mathrm{d}t} \,|\omega|^2 =
- \langle \omega | \Gamma' \omega \rangle \leq - \gamma'_0 \,|\omega|^2,
\end{equation}
where $\gamma'_0 > 0$ is the smallest eigenvalue of $\Gamma'$.
Since $|\omega|^2 \geq 0$, it follows from this equation that
$|\omega(t)|^2 \leq |\omega(0)|^2 \exp(-\gamma'_0 t)$ and,
consequently, $\lim_{t \to \infty} \omega(t) = 0$.
This leads to $\lim_{t \to \infty} \rho_{sf}(t) = 0$, because
$\rho_{sf}$ is a linear combination of operators
$| \omega \rangle \langle f |$ with $\omega \in \hs_s$, $f \in \hs_f$.
\item
Taking the trace of Eq.~(\ref{ss}), we obtain
\begin{equation}\label{trss}
\frac{\mathrm{d}}{\mathrm{d}t} \,\mathrm{Tr}\, \rho_{ss}(t) = -
\mathrm{Tr}\,\big(\Gamma \rho_{ss}(t)\big) \leq -
\gamma_0 \,\mathrm{Tr}\,\rho_{ss}(t).
\end{equation}
Here, $\gamma_0$ is the smallest eigenvalue of $\Gamma$.
From Eq.~(\ref{trss}) we derive the inequality
$\mathrm{Tr}\, \rho_{ss}(t) \leq \mathrm{Tr}\, \rho_{ss}(0)
\,\exp (-\gamma_0 t)$, whence we find
$\lim_{t \to \infty} \mathrm{Tr}\,\rho_{ss}(t) = 0$.
Since $\rho_{ss}(t)\geq 0$, the whole matrix must vanish
in the limit $t \to \infty$.
\item
Point ii) and $\mathrm{Tr} \,\rho(t) =
\mathrm{Tr}\, \rho_{ss}(t) + \mathrm{Tr}\, \rho_{ff}(t) = 1$ imply
$\lim_{t \to \infty} \mathrm{Tr}\, \rho_{ff}(t) = 1$.
\end{enumerate}

\paragraph{The simplest possible example:}
We choose the minimal dimensions $d_s = d_f = 1$,
the Hamiltonian
\begin{equation}\label{H-example}
\mathcal{H} = \left( \ba{cc} m & 0 \\ 0 & 0 \ea \right),
\end{equation}
and set $A_j=0$, i.e. we neglect any decoherence part.
Then, Eqs.~(\ref{ss}), (\ref{ff}) and
(\ref{rho-ff}) simplify considerably, providing the result
\begin{equation}\label{densitymatrix-explicit}
\rho(t) = \left( \ba{cc} e^{-\Gamma t} & 0 \\
0 & 1 - e^{-\Gamma t} \ea \right),
\end{equation}
where we have chosen the initial conditions
$\rho_{ss}(0) = 1$, $\rho_{ff}(0) = \rho_{sf}(0) = 0$.
This example illustrates the general fact that, for non-singluar $\Gamma$,
the probability loss of the system is
balanced by the probability increase of the decay states.

Considering the mixedness of the quantum states, we can use as measure
\begin{equation}
\delta(t) \equiv \mbox{Tr}\, \rho^2(t) =
1 - 2 \,e^{-\,\Gamma\,t} + 2 \,e^{-\,2\,\Gamma\,t}.
\end{equation}
We have $\delta(0) = 1$, $\lim_{t\to\infty} \delta(t) = 1$ and
$\delta(t) < 1$ for $0 < t < \infty$, i.e. in the beginning,
$\rho$ represents a pure state, then the state is mixed, whereas
for large times it approaches a pure state again,
when the system definitely has changed into the decay state.

The case of neutral kaons, where the Hamitonian $H$ and
the operator $B$ are $2\times 2$ matrices, has been investigated
in detail in Ref.~\cite{Caban-etal}.

\paragraph{Kraus operators:}
Finally, in the case of our example, we would like to consider
the time evolution of the density matrix as a dynamical map $V_t$.
In our case, $V_t$ can be represented by a
sum of operators~\cite{BreuerPetruccione,AlickiFannes}:
\begin{equation}\label{dynmap}
\rho(0) \;\longrightarrow\;  V_t[\rho(0)] =
\sum_j M_j(t) \,\rho(0) M_j^\dagger(t) = \rho(t),
\end{equation}
with the normalization $\sum_j M_j^\dagger(t) M_j(t) = \mathbbm{1}$
for the Kraus operators $M_j(t)$.
For the correspondence between the Lindblad and Kraus
operators see Ref.~\cite{Preskill-notes}.

The operator sum representation (\ref{dynmap}) is a very useful approach in quantum information to
describe the quantum operations or the specific quantum channels. In our example, the decay of a
particle corresponds to the amplitude damping channel of a quantum operation~\cite{NielsenChuang},
e.g., the spontaneous emission of a photon, and we can calculate the Kraus operators needed:
\begin{equation}
M_0(t) = \left( \ba{cc}
\sqrt{1-p(t)} & 0\\
0 & 1\\
\ea \right) \quad \mathrm{and} \quad
M_1(t) = \left( \ba{cc}
0 & 0\\
\sqrt{p(t)} & 0\\
\ea \right),
\end{equation}
with the probability $p(t) = 1 - e^{-\Gamma t}\,$.

\paragraph{Conclusions:}
In this letter we have considered a time evolution given by
Eq.~(\ref{Lindbladequation}),
with a non-hermitian Hamiltonian $H_\mathrm{eff}$ and a dissipator of the
Lindblad form~(\ref{dissipator}).
Assuming that the non-hermitian part of $H_\mathrm{eff}$ describes
``particle decay,'' we have shown that
such a time evolution is completely positive.
Our strategy was to add the space of ``decay states'' $\hs_f$ to the space of
states $\hs_s$ and to extend the time evolution in a straightforward way to
the full space $\hs_s \oplus \hs_f$, such that this full time evolution is
probability--conserving.
With the initial condition that the density matrix is only non-zero on
$\hs_s$, the Lindblad operator $\mathcal{B}$ of Eq.~(\ref{operator-B}),
which is responsible for the decay,
shifts states from $\hs_s$ to $\hs_f$, whereas at
the same time the $\mathcal{A}_j$
terms cause decoherence and dissipation on $\hs_s$.
Thus, particle decay and decoherence/dissipation are related phenomena,
which can be described by the same
formalism of a completely positive time evolution.

\vspace{8mm}

\noindent
\textit{Acknowledgement:}
The authors acknowledge financial support of EURIDICE
HPRN-CT-2002-00311.


\newpage

\begin{thebibliography}{99}

\bibitem{BreuerPetruccione}
H.-P. Breuer and F. Petruccione,
\textit{The theory of open quantum systems},
Oxford University Press 2002.

\bibitem{AlickiFannes}
R. Alicki and M. Fannes,
\textit{Quantum dynamical systems},
Oxford University Press, 2001.

\bibitem{Joos}
E. Joos, \textit{Decoherence through interaction with the environment},
in: \textit{Decoherence and the appearance of a classical world in quantum
  theory},
D. Giulini et al. (eds.), Springer Verlag, Heidelberg, 1996, p.35.

\bibitem{KublerZeh}
O. K\"ubler and H.D. Zeh,
Ann. Phys. (N.Y.) \textbf{76}, 405 (1973).

\bibitem{JoosZeh}
E. Joos and H.D. Zeh, Z. Phys. B \textbf{59}, 223 (1985).

\bibitem{Zurek}
W.H. Zurek, Physics Today \textbf{44}, 36 (1991).

\bibitem{EllisHagelinNS}
J. Ellis, J.S. Hagelin, D.V. Nanopoulos, and M. Srednicki,
Nucl. Phys. B \textbf{241}, 381 (1984).

\bibitem{HuetPeskin}
P. Huet and M.E. Peskin,
Nucl. Phys. B \textbf{434}, 3 (1995)
[hep-ph/9403257].

\bibitem{EllisLopezMN}
J. Ellis, J.L. Lopez, N.E. Mavromatos, and D.V. Nanopoulos,
Phys. Rev. D \textbf{53}, 3846 (1996)
[hep-ph/9505340].

\bibitem{BertlmannDurstbergerHiesmayr2002}
R.A. Bertlmann, K. Durstberger, and B.C. Hiesmayr,
Phys. Rev. A \textbf{68}, 012111 (2003)
[quant-ph/0209017].

\bibitem{BG3}
R.A. Bertlmann and W. Grimus,
Phys. Rev. D \textbf{64}, 056004 (2001)
[hep-ph/0101160].

\bibitem{BernabeuMavromatosP}
J. Bernab\'eu, N.E. Mavromatos, and J. Papavassiliou,
Phys. Rev. Lett. \textbf{92}, 131601 (2004)
[hep-ph/0310180].

\bibitem{BertlmannSchladming}
R.A. Bertlmann, \textit{Entanglement, {Bell} inequalities and decoherence in
  particle physics},
Lecture Notes in Physics, Springer-Verlag, Berlin, 2005,
[quant-ph/0410028].

\bibitem{Lindblad}
G. Lindblad,
Comm. Math. Phys. \textbf{48}, 119 (1976).

\bibitem{GoriniKossakowskiSudarshan}
V. Gorini, A. Kossakowski, and E.C.G. Sudarshan,
J. Math. Phys. \textbf{17}, 821 (1976).

\bibitem{Preskill-notes}
J. Preskill, \textit{Lecture notes},
http://theory.caltech.edu/people/preskill/ph229/.

\bibitem{Caban-etal}
P. Caban, J. Rembieli\'nski, K.A. Smoli\'nski, and Z. Walczak,
Phys. Rev. A \textbf{72}, 032106 (2005)
[quant-ph/0506183].

\bibitem{QuangPham}
Quang Ho-Kim and Pham Xuan-Yem,
\textit{Elementary particles and their interactions},
Springer Verlag, Berlin, 1998.

\bibitem{NielsenChuang}
M.A. Nielsen and I.L. Chuang,
\textit{Quantum computation and quantum information},
Cambridge University Press, Cambridge, 2000.

\end{thebibliography}
\end{document}